\documentstyle [12pt,aaspp4]{article}
\def\beq{\begin{equation}}
\def\eeq{\end{equation}}
\def\bey{\begin{eqnarray}}
\def\eey{\end{eqnarray}}
\def\pppm{\rm P^3M}

\def\mpc{\,h^{-1}{\rm {Mpc}}}
\def\kms{\,{\rm {km\, s^{-1}}}}

\def\gs{\mathrel{\raise1.16pt\hbox{$>$}\kern-7.0pt
\lower3.06pt\hbox{{$\scriptstyle \sim$}}}}
\def\ls{\mathrel{\raise1.16pt\hbox{$<$}\kern-7.0pt
\lower3.06pt\hbox{{$\scriptstyle \sim$}}}}
\def\gtsima{$\; \buildrel > \over \sim \;$}
\def\ltsima{$\; \buildrel < \over \sim \;$}
\def\prosima{$\; \buildrel \propto \over \sim \;$}
\def\gsim{\lower.5ex\hbox{\gtsima}}
\def\lsim{\lower.5ex\hbox{\ltsima}}
\def\simgt{\lower.5ex\hbox{\gtsima}}
\def\simlt{\lower.5ex\hbox{\ltsima}}
\def\simpr{\lower.5ex\hbox{\prosima}}
\def\la{\lsim}
\def\ga{\gsim}

\begin{document}
\title {Spatial correlation function and pairwise velocity dispersion 
of galaxies: CDM models versus the Las Campanas Survey }
\author {Y.P. Jing$^{1,2}$, H.J. Mo$^{2}$, G. B\"orner$^2$} 
\affil{$ ^1$Research Center for the Early Universe,
School of Science, University of Tokyo, Bunkyo-ku, Tokyo 113, Japan}
\affil {$ ^2$Max-Planck-Institut f\"ur Astrophysik,
Karl-Schwarzschild-Strasse 1, 85748 Garching, Germany}

\received{---------------}
\accepted{---------------}

\begin{abstract}
We show, with the help of large N-body simulations, that the
real-space two-point correlation function and pairwise velocity
dispersion of galaxies can both be measured reliably from the Las
Campanas Redshift Survey.  The real-space correlation function is well
fitted by the power law $\xi(r)=(r_0/r)^\gamma$ with $r_0=(5.06\pm
0.12)\mpc$ and $\gamma=1.862\pm 0.034$, and the pairwise velocity
dispersion at $1\mpc$ is $(570\pm 80)\kms$. A detailed comparison
between these observational results and the predictions of current CDM
cosmogonies is carried out. We construct 60 mock samples for each
theoretical model from a large set of high resolution N-body
simulations, which allows us to include various observational
selection effects in the analyses and to use exactly the same methods
for both real and theoretical samples. We demonstrate that such a
procedure is essential in the comparison between models and
observations.

The observed two-point correlation function is significantly flatter
than the mass correlation function in current CDM models on scales
$\la 1\mpc$. The observed pairwise velocity dispersion is also lower
than that of dark matter particles in these models.  We propose a
simple antibias model to explain these discrepancies.  This model assumes
that the number of galaxies per unit dark matter mass, $N/M$,
decreases with the mass of dark haloes.  The predictions of CDM models
with $\sigma_8\Omega_0^{0.6}\sim 0.4$-0.5 and $\Gamma \sim 0.2$ are in
agreement with the observational results, if the trend of $N/M$ with
$M$ is at the level already observed for rich clusters of
galaxies. Thus CDM models with $\Gamma \sim 0.2$ and
with cluster-abundance normalization are
consistent with the observed correlation function and pairwise
velocity dispersion of galaxies.  A high level of velocity bias is not
required in these models.
\end{abstract}


\section {INTRODUCTION}

Our understanding of the large-scale structure in the Universe
comes mainly from large surveys of galaxies. Given a well-defined
galaxy sample, we can derive statistical measures of the 
large scale structure and compare them with model predictions.
The two-point correlation function (hereafter TPCF) of galaxies is
such a statistic that can be derived easily from
a galaxy sample. Model predictions for this statistic by current 
cold dark matter (CDM) models can also be made either from
high-resolution N-body simulations or from empirical fitting formulae
calibrated by such simulations. As a result, the TPCF of galaxies 
has long been used as an important diagnostic for distinguishing
theoretical models (e.g. Peebles 1980). Comparing model predictions
for the mass correlation with the angular two-point correlation 
function of galaxies in the APM survey 
(e.g. Maddox et al. 1990, 1996; Baugh 1996), Peacock (1997) 
and Jenkins et al. (1996) conclude that current CDM models are 
inconsistent with the observational result unless there is a 
scale-dependent bias in the distribution of galaxies 
relative to that of the mass. 
This is obviously an important result, but clearly new observational 
results and objective comparisons between models and observations are
needed to quantify the significance of such an inconsistency.

The pairwise velocity dispersion (hereafter PVD) 
of galaxies is another important quantity 
which can be measured from redshift surveys (Peebles 1980; Suto 1993). 
As a measure of the relative motion of galaxies, this
quantity probes the mean mass density of the universe and
the clustering power on small scales, and so has also been 
widely used as a critical test for cosmogonic 
models (e.g. Davis et al. 1985). However, as pointed out by
Mo, Jing, \& B\"orner (1993) based
on their analysis of various redshift surveys then available, 
the early result of Davis \& Peebles (1983; see also
Beans et al. 1983), which gives 
a value of about $340\kms$ for the PVD, may be biased towards
low values, and
the average value is more likely to be about $400$-$600 \kms$.
They also emphasized that the value of the PVD is very sensitive 
to the presence (or absence) of rich clusters in a sample  
and the surveys then available are too small to give a fair
estimate. Similar conclusions have since been reached by many other
authors based on various redshift samples (Zurek et al. 1994;
Fisher et al. 1994; Guzzo et al. 1995; Marzke et al. 1995; 
Somerville, Davis \& Primack 1996; Ratcliffe et al. 1997).
As a result, the constraint on theoretical models given by the
observed PVD is still uncertain. As demonstrated clearly 
in Mo, Jing \& B\"orner (1997), to have the PVD fairly sampled, 
one needs samples that contain many rich clusters. All
galaxy redshift samples used in the analyses mentioned above
are too small to qualify for this purpose, and so a much
larger galaxy sample is needed. 

  In this paper we show, with the help of large N-body 
simulations, that the spatial TPCF and the PVD of galaxies can now
both be measured accurately from the Las Campanas Redshift Survey  
(Shectman et al. 1996, hereafter LCRS), the largest 
redshift sample so far available. This enables us to carry out
a detailed comparison
between the observed results and the predictions of current
CDM cosmogonies. We construct 60 mock samples for each 
theoretical model from a large set  of high resolution N-body 
simulations. Consequently we are able to include various 
observational selection effects in the analyses and to quantify 
the statistical significance of the observational results. 
The mock catalogues also allow exactly the same methods to be applied 
both to the real and to the theoretical samples, making it possible
for us to carry out an objective comparison between  
theoretical predictions and the observational results.

 It is important to point out that a large set of mock samples is
{\it essential} in the comparison between models and observations.
This is particularly true for the PVD. First, as discussed above,
the PVD is very sensitive to the presence (or absence)
of rich clusters in a sample. Such a sampling effect is
difficult to model analytically and needs to be quantified
by mock samples. Second, the observed PVD is estimated from the
redshift distortion of the two-point correlation
function and so depends both on the distribution
function of peculiar velocities and on the mean infall 
velocities of galaxy pairs. Both the distribution
function and the mean infall are not known 
{\it a priori}. Mock samples can help to quantify the 
importance of these uncertain factors. Finally, the observed PVD 
is an average of the {\it true} PVD along the line-of-sight, 
and the relation between the two is difficult to quantify analytically
because the true PVD is a complicated function of the
separation of galaxy pairs in real space. 
Such systematics can only be taken into account
when mock samples are analyzed in the same way 
as the real sample. 

 The spatial TPCF and PVD of galaxies in the LCRS
have already been estimated by the survey group (Lin et al. 1997). 
Our results are generally in agreement with theirs, although
our analysis differs in many ways from theirs.
As discussed above, one main purpose of this paper is to use 
these observational results to constrain theoretical models.
We will therefore adopt our own results for comparison
with the model predictions, because we can then apply the same method 
to the mock catalogues. In addition we will carefully quantify 
the effects of the ``fiber collision'' limitation on the two statistics; 
this complements the work of Lin et al. (1997). 
Our work is also distinct from that of Peacock (1997) and Jenkins et
al. (1996), because we use the spatial correlation function
measured directly from the LCRS instead of the one deconvolved
from the angular correlation function of the APM galaxies,
and because we use both the TPCF and PVD to constrain models. 

 As we will show below, the observed TPCF of 
the LCRS galaxies is significantly flatter 
than the mass correlation function in (some) 
current CDM models
on scales $\la 1\mpc$. This is consistent with the results
based on the correlation function of APM galaxies. 
The observed PVD of galaxies on small scales is also 
lower than the PVD of dark matter particles in these models. 
Thus, unless galaxies are biased with respect to the mass, 
these models will all be ruled out. However, as discussed by 
Dekel \& Rees (1987), many physical processes in galaxy 
formation can give rise to biases in the distribution 
of galaxies relative to the underlying mass density field.
A velocity bias in the sense that galaxies have systematically 
lower peculiar velocities than dark matter particles 
has also been found in numerical simulations (e.g. Carlberg \&
Couchman 1989; Katz et al. 1992; Cen \& Ostriker 1992; 
Frenk et al. 1996). Unfortunately, detailed relationship
between galaxy and mass distributions is still unknown,
and all physical models for the density and velocity biases are 
uncertain. Given this situation, we think it is more useful
to use a simple but plausible phenomenological model 
to gain some insight into the problem. Specifically, we will 
assume the number of galaxies per unit dark matter
mass, $N/M$, is lower in massive haloes than in less massive ones. 
The main motivation for this assumption comes from the fact that
such a trend has indeed been observed in the CNOC sample
of rich clusters (Carlberg et al. 1996) and this
hypothesis can easily be tested 
further. As we will see, this simple model does bring the 
predictions of (some) CDM models in agreement with the observational 
results, if we assume a trend of $N/M$ with $M$ at the observed level.
This implies that such models are still consistent with observations.

 The arrangement of the paper is as follows. We describe the
observational samples and mock catalogues in Section 2.
The statistical results for the TPCF and PVD for both the real and 
mock samples are presented and compared in Section 3. The
phenomenological bias model and its effects on both TPCF
and PVD are discussed in Section 4. Finally, in Section 5 we
present a brief discussion of our results. For completeness, 
two appendices are included where we investigate in detail 
the `fiber collision' effect and examine
whether an unbiased estimate of the 
TPCF and PVD of galaxies can be obtained from a sample 
like the LCRS.
 
\section {OBSERVATIONAL SAMPLES AND MOCK CATALOGUES}

We use the Las Campanas Redshift Survey (Shectman et al. 1996; LCRS)
to determine the spatial two-point corelation function and the
pairwise velocity dispersion. This survey is the largest redshift
survey, which is now publicly available. Our main sample consists of
all galaxies with recession velocities between 10,000 and 45,000
$\kms$ and with absolute magnitudes (in the LCRS hybrid R band)
between $-18.0$ and $-23.0$. There are 19558 galaxies in this sample,
of which 9480 are in the three north slices and the rest in the three
south slices.  Analyses are carried out for the full sample, as well
as separately for the north and south subsamples.

The LCRS is a well-calibrated sample of galaxies, ideally suited for
statistical studies of large-scale structure. All known systematic
effects in the survey are well quantified and documented (Shectman et
al. 1996; Lin et al. 1996), and so most can be corrected easily in
statistical analyses. The only exception is the `fiber collision'
limitation which prevents two galaxies in one $\sim 1.5\times 1.5\,
{\rm deg}^2$ field from being observed when they are closer than
$55''$ on the sky, because it is impossible to put fibers on both
objects simultaneously.  Here we will use extensively mock catalogues
generated from N-body simulations to quantify this effect. When
comparing models with observations, we will use statistical results
measured from mock samples that take into account of this fiber
collision effect. We make corrections to the observed results with the
help of a comparison between mock samples in which the fiber collision
is included and excluded.

 The N-body simulations used here are generated with our $\pppm$ code
(Jing \& Fang 1994) with $128^3$ particles.  We consider three
spatially-flat cosmological models with $\Omega_0+\lambda_0=1$, where
$\Omega_0$ is the density parameter and $\lambda_0$ the cosmological
constant.  The linear perturbation spectra are assumed to be of
CDM-type as given in Bardeen et al. (1986), fixed by the shape
parameter\footnote{The $\Gamma$ parameter was introduced by White et
al (1993) for the transfer function of Davis et al. (1985). 
That transfer function differs slightly from what is used here.
Thus our linear power spectrum
differs (slightly) from that of White et al. (1993) 
even for the same value of $\Gamma$.} 
$\Gamma\equiv \Omega_0 h$ (where $h$ is the Hubble constant in
units of $100\kms {\rm Mpc}^{-1}$) and the normalization $\sigma_8$
(which is the present rms of the density contrast in top-hat windows
of radius $8\mpc$ given by the linear power spectrum).  Each model is
then specified by the three parameters: $\Omega_0$, $\Gamma$ and
$\sigma_8$. The three models considered here have ($\Omega_0$,
$\Gamma$, $\sigma_8$)=(0.2,0.2,1), (0.3,0.2,1) and (1,0.5,0.62)
respectively. The parameters chosen for the $\Omega_0=1$ model are
similar to those for the standard CDM model, while the two low-$\Omega$
models are chosen because they are compatible with most observational
constraints.  For the standard CDM model, the simulation box is
$300\mpc$ on each side, the force resolution is $0.12\mpc$, and 400
time steps are used to evolve the simulations.  For the two
low-$\Omega$ models, these parameters are $256\mpc$, $0.1\mpc$ and
585, respectively.  Six realizations are generated for each model.
Since the depth and width of the LCRS slices are larger than the box
size, we have to replicate the simulations periodically along each
axis. This should have little effect on the statistical results if the
mock slices avoid the principal planes and if the scales of interests
are much smaller than the box size.  The resolutions of these
simulations are sufficiently high for the purpose of this paper, as
shown clearly by the tests presented in Appendix A.

To generate a mock catalogue, we first select a random position
in the simulation box as the origin (i.e. the position of the
`observer').  Assuming one of the three axes to point towards the north
pole, we generate a photometric catalog according to the angular
boundaries (Shectman et al. 1996), the luminosity function (Lin et al.
1996) and the limiting magnitudes of each 
$\sim 1.5\times 1.5 \,{\rm deg^2}$ field of the survey.  
We choose randomly 101 `galaxies' from each of the 112-fiber fields
and 45 from each of the 50-fiber fields if there are more
`galaxies' in the field. Otherwise all `galaxies' in the field
are chosen.  The number of `galaxies' in each field is slightly 
less than the number of the available fibers because in the
observation about $10$ percent of the observed spectra turned out to 
be star spectra or could not be identified.
The fiber collision effect is simulated by eliminating 
one galaxy from any pair (in one field) that is closer than $55''$ on
the sky. The results of such eliminations are sometimes not unique,
and we have adopted an algorithm which minimizes the number of 
the eliminated `galaxies'.  Finally, each `galaxy' is
assigned a probability of being observed 
(or the observed fraction), in the same way as
Lin et al. (1996) did for the real observation. 
Ten mock samples are generated for each simulation, 
and so 60 mock samples are used for each model. Since the volume of
the simulation box is about 10 times as large as than the effective volume of
the LCRS, mock samples thus obtained are approximately
independent of each other.

\section{COMPARISON BETWEEN MODELS AND OBSERVATIONS}

\subsection{Correlation function and pairwise velocity dispersion 
from the LCRS}

We estimate the redshift-space two-point correlation $\xi_z(r_p,\pi)$ by
\beq \label{xirppi}
\xi_z(r_p,\pi)={4RR(r_p,\pi)\cdot DD(r_p,\pi)\over
 [DR(r_p,\pi)]^2}-1,
\eeq
where $DD(r_p,\pi)$ is the count of (distinct) galaxy-galaxy pairs 
with perpendicular separations in the bin $r_p\pm 0.5\Delta r_p$ and 
with radial separations in the bin $\pi\pm 0.5\Delta \pi$,
$RR(r_p,\pi)$ and $DR(r_p,\pi)$ are similar counts of pairs
formed by two random points and by one galaxy and one random point,
respectively (e.g. Hamilton 1993). In computing the pair counts, 
each galaxy is weighted by the inverse of the observed fraction 
(see Lin et al. 1996) to correct for the sampling effect and
for the apparent-magnitude and surface-brightness
incompleteness. The random sample, which contains 100,000 points, 
is generated in the same way as the mock samples,  except that 
the points are originally randomly distributed in space.  
The projected two-point correlation function $w(r_p)$ 
is estimated from
\beq \label{wrp}
w(r_p)=\int_0^\infty \xi_z(r_p,\pi) d\pi =\sum_i
\xi_z(r_p,\pi_i)\Delta \pi_i,
\eeq
where $\xi_z(r_p,\pi_i)$ is measured by equation (\ref{xirppi}).
The summation runs from $\pi_1=0.5\mpc$ up to $\pi_{50}=49.5\mpc$ with
$\Delta \pi_i=1\mpc$. The resulted $w(r_p)$ is, however, quite robust to
reasonable changes of the upper limit of $\pi_i$; an upper limit
$30\mpc$ for $\pi_i$ does not make any notable difference.  
Note also that our definition of $w(r_p)$ differs from
that of Davis \& Peebles (1983) by a factor 2, 
as ours assumes $\pi>0$.
We have tested the above procedure by applying it 
to the mock samples. The results of the test are presented in
Appendix A. It is shown there that this procedure gives 
an unbiased estimate of $w(r_p)$ if there is no fiber 
collision effect.
Fiber collisions suppress $w(r_p)$ 
on very small scales, as expected. 
However, as shown in Fig. 6, the suppression is
generally small, amounting to only about $14\%$ in $w(r_p)$ at
$r_p=0.1\mpc$, and dying off very quickly as the scale increases. 
This suppression effect can easily be corrected because it is
systematic and depends only weakly on the intrinsic 
clustering power.  Such a correction is not even necessary in our 
comparison between model predictions and the observational
results for $w(r_p)$, because the fiber collision effects are already  
included in the mock samples. Unless explicitly
stated, results quoted for both the mock and the
observational samples have not been corrected for the suppression due to 
bfiber collisions.

The projected two-point correlation function for the LCRS survey is
presented in Figure 1. Triangles are the results for the whole sample.
Error bars are estimated by the bootstrap resampling technique (Barrow
et al. 1984). We generate 100 bootstrap samples and compute $w(r_p)$
for each sample using the weighting scheme (but not the approximate
formula) given in Mo, Jing \& B\"orner (1992). The error bars are the
scatter of $w(r_p)$ among these bootstrap samples. Our test on
mock samples shows that the bootstrap errors are comparable (within a
factor 2) to the standard scatter among different mock samples, and so
it does not matter much which error estimate is used in our
discussion. A power-law fit of the two-point correlation function
$\xi$ to the observed $w(r_p)$ over $r_p<28\mpc$ yields 
\beq\label{xi}
\xi(r)=(r_0/r)^\gamma, \eeq 
with 
\beq \label{r0gamma}
r_0=5.01\pm0.05\mpc;\hskip 1.2cm \gamma=1.825\pm 0.018.  \eeq 
As discussed in Appendix B, fiber collisions suppress the correlation
function on small scales by a small amount.  Such a suppression is
systematic and can easily be corrected with the help of mock
samples. The error bars given by the bootstrap method may be
underestimated by a factor of 2, as shown in section 3.2.  The values 
of $r_0$ and $\gamma$ after these corrections are
\beq \label{r0gammax} 
r_0=5.06\pm0.12\mpc;\hskip 1.2cm \gamma=1.862\pm
0.034.  \eeq 
The fitting result given by equation (\ref{r0gamma}) is
shown in Fig. 1 and it is clear that the observational data are well
fitted by the power law.  We have also analysed the north and south
subsamples separately, and the results are also depicted in
Fig. 1. The two results agree with each other reasonably well,
especially for $r_p<5\mpc$. The error bars for the subsamples are
about 1.5 times as large as those of the whole sample, and so all the
results are consistent with each other. The real-space correlation
function derived here is significantly steeper than that derived from
the angular correlation function of the APM survey (e.g. Maddox et
al. 1990; Baugh 1996).
At the moment it is not clear whether this is due to some unknown
systematics in the two surveys, or due
to the fact that galaxies in the APM survey are selected in a bluer
band than the LCRS. If the latter is the reason, then the difference 
in the slope will have interesting implications for the theories of
galaxy formation. 

The pairwise velocity dispersion of galaxies is measured by modeling 
the redshift distortion in the observed redshift-space correlation function
$\xi_z(r_p,\pi)$. The relation between $\xi_z(r_p,\pi)$ and the 
real-space correlation function $\xi(r)$ is usually assumed to be 
\beq\label{xizmodel}
1+\xi_z(r_p,\pi)=\int
f(v_{12})\left[1+\xi(\sqrt{r_p^2+(\pi-v_{12})^2})
\right]dv_{12},
\eeq 
where $f(v_{12})$ is the distribution function of the relative
velocity (of galaxy pairs) along the line-of-sight 
(see e.g. Fisher et al. 1994). Based on observational
(Davis \& Peebles 1993; Fisher et al. 1994) and 
theoretical considerations (e.g. Diaferio \& Geller 1996; Sheth 1996), 
an exponential form is usually adopted for $f(v_{12})$:
\beq \label{fv12}
f(v_{12})={1\over \sqrt{2}\sigma_{12}} \exp \left(-{\sqrt{2}\over
\sigma_{12}} \left| v_{12}-\overline{v_{12}}\right| \right),
\eeq
where $\overline {v_{12}}$ is the mean and 
$\sigma_{12}$ the dispersion of the 1-D pairwise peculiar
velocities.

 Assuming an infall model for $\overline v_{12}(r)$ and modelling
$\xi(r)$ from the projected correlation function, one can estimate the
pairwise velocity dispersion $\sigma_{12}$ by comparing the observed
redshift-space correlation function, $\xi^{\rm obs}_z(r_p,\pi)$, 
with the modelled one, $\xi^{\rm mod}_z(r_p,\pi)$,
given by the right-hand-side of equation (\ref{xizmodel}). 
In practice we estimate $\sigma_{12}$ by minimizing
\beq\label{minimization}
\min \left\{\sum_i\left[ {\xi^{\rm obs}_z(r_p,\pi_i)-
\xi^{\rm mod}_z(r_p,\pi_i)\over \sigma_{\xi_z}^{\rm obs}
(r_p,\pi_i)}
\right]^2\right\},
\eeq
where the summation is over all $\pi$ bins for a fixed $r_p$ and
so $\sigma_{12}$ is generally a function of $r_p$,
$\sigma_{\xi_z}^{\rm obs}(r_p,\pi_i)$ is the error of $\xi^{\rm
  obs}_z(r_p,\pi_i)$ estimated by the bootstrap method.

It should be realized that the PVD measured with the above procedure 
is not the same as the one given directly by the peculiar velocities 
of galaxies. In the reconstruction of $\sigma_{12}$ from 
equations (\ref{xizmodel})-(\ref{minimization}),
the infall velocity $\overline v_{12}(r)$ 
is not known {\it a priori } for the observed galaxies. Neither are
the forms of $f(v_{12})$ and $\xi(r)$. 
The models used for these functions are therefore only approximate. 
Furthermore, the PVD estimated from the redshift distortion 
is a kind of average of the true PVD along the line of sight, and
since the true PVD depends on the separations of galaxy pairs 
in {\it real} space (see Appendix A), the two quantities are different
by definition.
Unfortunately, at the moment we do not have a better
method to get rid of these problems. As we show in Appendix A
using our simulation results, the bias caused by these systematics
may be significant, although the PVD estimated from the redshift 
distortion still measures the true PVD in some (complicated) way. 
Thus, when we compare model predictions with the observational results,
these systematics must be treated carefully. The most objective 
approach is to construct a large set of mock samples
from the theoretical models and to analyze them in the same way 
as the real samples. This we do in Section 3.2.

 Before going to Section 3.2, let us present our estimate of
$\sigma_{12}(r_p)$ for the LCRS. Here we assume 
an infall model based on the self-similar solution:
\beq\label{infall}
\overline{v_{12}}({\bf r})= -{y\over 1+(r/r_{\star})^2},
\eeq 
where $r_{\star}=5\mpc$ and $y$ is the radial separation in the real space. The reason for adopting this assumption
is that this infall model has been widely used in
previous analyses and is a good approximation to the real 
infall pattern in CDM models with $\sigma_8 \Omega_0^{0.6}\approx
0.5$. The results are presented in Figure 2.
The PVD of the LCRS galaxies is about $(550\pm 50)\kms$ at
$r_p=1\mpc$. As discussed in Appendix B and section 3.2, the fiber
collision effect reduces the PVD by about $20\kms$, and the bootstrap
error of the PVD is about 30 percent smaller than the error given by mock
samples. Thus our best estimate of $\sigma_{12}(1\mpc)$ is $(570\pm
80)\kms$.  The results for the northern and southern subsamples are
very similar on scales $r_p=0.2$-$5.0\mpc$, implying that a fair
estimate of the small-scale PVD of galaxies can be obtained from
galaxy surveys as large as the LCRS. The LCRS contains about 30
clusters of galaxies and so it samples the mass function of clusters
reasonably well. As discussed in Mo, Jing \& B\"orner (1997), such a
data set is needed to have the PVD of galaxies fairly sampled.

\subsection{Comparison with model predictions}

Having shown that both the TPCF and PVD of galaxies can be estimated
reliably from the LCRS, we now use our results to constrain 
theoretical models. To do this,  we apply exactly the same statistical
procedure used for the real samples to mock samples derived from the 
N-body simulations. The projected TPCF and PVD of dark matter
particles are estimated for each mock sample. The averages 
of these two quantities and the $1\sigma$ scatter among the mock samples
are plotted in Figure 3 for the three theoretical models. 
The results for the LCRS are also included
for comparison. 

The $w(r_p)$ predicted by the  
two low-$\Omega$ models with $\sigma_8=1$ are in good agreement 
with the observed one on scales larger than $\sim 5\mpc$. 
On smaller scales, however, the predictions of both models 
lie above the observational result. 
Although the TPCF we obtain here 
from the LCRS is steeper than that derived from the angular
correlation function of the APM survey, our conclusion about
the shape of the two-point correlation function in these two
models is qualitatively the same as that reached by 
Efstathiou et al. (1990) and re-stressed by Peacock (1997)
and Klypin et al. (1996) based on the APM result. The projected 
TPCF predicted by the
standard CDM model is lower than that of the galaxies, because 
of the lower normalization, $\sigma_8=0.62$, in this model. 
If we shift the model prediction upwards by a factor of 
$1/0.62^2\approx 2.6$, as implied by a linear biasing factor
$b=1/\sigma_8=1.6$,
the model prediction fits the observed $w(r_p)$ on intermediate 
scales. The discrepancy on scales $r_p\ga 5\mpc$ is due to the 
well-known fact that this model does not have large enough
power on large scales. It is also apparent from the figure that
this model predicts too steep a $w(r_p)$ around $r_p=1\mpc$.
A formal $\chi^2$ test shows that the discrepancy between
the model predictions and the observational result is
highly significant. 
Thus, a scale-dependent bias is required by all three models
in order for them to be compatible with the observed real-space 
correlation function given by the LCRS.

As shown clearly in Figure 3, the PVDs of dark matter particles
predicted by all three models are higher than the observed value
for $r_p<5\mpc$. On larger separations, the statistical fluctuations
become very large and the result is very sensitive to the infall model
adopted (see below). We have tried to use a $\chi^2$ test to quantify the
discrepancy between the model predictions and the observational
result. However, the distribution of the PVDs given by mock samples 
differs substantially from a Gaussian, and we have to 
use a different measure to quantify the discrepancy. 
Since the PVD is tightly correlated
among different $r_p$ bins, we define the probability
$P(<\sigma_{12}^{\rm obs})$ that the PVD given by a mock sample is lower
than the observational result in some range of $r_p$. Obviously 
$P(<\sigma_{12}^{\rm obs})$ characterizes the difference between the
model prediction and the observational result. Using the mean value of
PVD in the three $r_p$ bins around $r_p=1\mpc$, we find,
based on 60 mock samples, that  
$P(<\sigma_{12}^{\rm obs})=$1/60, 0/60 and 0/60 for the
$\Omega_0=0.2$, $0.3$, and $1.0$ models, respectively.
Using the value of PVD in one bin near
$r_p\approx 1\mpc$ gives the same result. Thus the PVDs 
predicted by all three models are significantly
higher than the observed value.

 The PVD for the mass on small scales is proportional
to $\sigma_8\Omega_0^{0.5}$ (Mo, Jing \& B\"orner 1997).
From the observed abundance of galaxy clusters, it has been 
argued that $\sigma_8\Omega_0^{0.6}$ is about $0.5$ 
(White et al. 1993; see also Viana \& Liddle 1996, Eke et al. 1996;
Kitayama \& Suto 1997).  
Unfortunately the observational results are still uncertain.
For example, lower values, with $\sigma_8\Omega_0^{0.6}\sim 0.4$,
may still be consistent with observations
(Carlberg et al. 1996; Bahcall et al. 1997). The values adopted for the 
$\Omega_0=0.2$ model (which has $\sigma_8\Omega_0^{0.6}\approx 0.38$) 
and for the $\Omega_0=0.3$ model (which has $\sigma_8\Omega_0^{0.6}
\approx 0.5$) are consistent with these observational results.
The value of $\sigma_8\Omega_0^{0.6}$ 
taken for the $\Omega_0=1$ model ($\approx 0.6$) is a little higher than
the observed one. If we lower $\sigma_8\Omega^{0.6}$
in this model to 0.5, the model prediction is still 
higher than the observational result. Thus, unless the PVD
of galaxies is, for some reasons, biased low relative to that 
of the mass, all the three models will have problems to 
match the observational result.  

  As discussed in Section 3.1, there is no compelling reason for 
assuming equation (\ref{infall}) as the infall model for 
the LCRS galaxies. To check the effect of this assumption, 
we also use the infall pattern derived directly from the 
CDM simulations. The dotted lines in the right panels of Fig.3 show
the results given by such an infall model. 
The open circles in each panel
are the result for the LCRS sample analysed using the infall
model given by the CDM model shown in the same panel. 
As one can see, all results are qualitatively the same 
as those given by the self-similar infall model (equation
\ref{infall}). Hence our results are not sensitive to the 
changes in the infall model.

\section{A PHENOMENOLOGICAL BIAS MODEL FOR $w(r_p)$ AND
$\sigma_{12}$}
 
Our comparison between the predictions of current CDM models and the
observational results from the LCRS show that the galaxy distribution
must be biased relative to the underlying mass for these models to be
viable. Since the mechanisms for such biases are still unclear, we use
a simple but plausible phenomenological model to gain some insight
into the problem. Specifically, we will assume the number of galaxies
per unit dark matter mass, $N/M$, is lower in more massive haloes. As
will be discussed in Section 5, the motivation for this assumption
comes from the fact that such a trend has been observed for clusters
of galaxies (Carlberg et al. 1996).  We will take a simple power-law
form for the dependence of $N/M$ on $M$: \beq\label{ntom} N/M\propto
M^{-\alpha}, \eeq where $M$ is the cluster mass, $N$ is the number of
`galaxies' in the cluster, and $\alpha$ is the parameter describing
the dependence.  In our discussion here, clusters are defined in the
N-body simulations by the friends-of-friends method with the linkage
parameter equal to 0.2 times the mean separation of particles.  Since
the predicted TPCF is steeper and the PVD is higher than the observed
values on scale $\sim 1\mpc$, the observational results require
$\alpha$ to be positive, namely, there are fewer galaxies per unit
dark matter mass in massive clusters than in poorer clusters.

To incorporate this model into our correlation analysis, we just give
each `mock' galaxy a weight which is proportional to $M^{-\alpha}$,
where $M$ is the mass of the cluster in which the `galaxy' resides.
Since the mass of individual particles in our simulations is
approximately that of galactic halos, this procedure implies
that we use equation (\ref{ntom}) for all dark halos 
more massive than those of typical galaxies. Our results for the
TPCF and PVD do not change significantly if equation (\ref{ntom})
is applied only for halos at least ten times more massive,
because the effects
arise mainly from the most massive halos.   
We have run a few trials for different values of $\alpha$. Figure
4 shows the results for $\alpha=0.08$. With this value, the
projected TPCFs predicted by the two low-$\Omega$ models
fits the observed one very well both in shape and in amplitude.
The prediction of the $\Omega_0=1$ model is also 
consistent with the observational result on scales less than 
$5\mpc$ if the mass correlation function is boosted by the 
linear bias factor $b=1/\sigma_8$. On larger scales,
this model does not have high enough clustering power to
match the observation, as is known from other observations.

  The agreement between the model predictions and the observational
results in the PVD is also improved substantially for all three 
models.  The probability $P(<\sigma_{12}^{\rm obs}$) now
increases to 10/60 for the $\Omega_0=0.2$ model and to 1/60 
for the other two models. In the two low-$\Omega$ models, 
the shapes of the predicted $\sigma_{12}(r_p)$ are similar to the
observed one. The amplitude of PVD predicted by the $\Omega_0=0.2$
model is consistent with the observed value, while that
predicted by the $\Omega_0=0.3$ model is about 30 percent too high.
Such a difference between these two models is expected,   
because the value of $\sigma_8\Omega_0^{0.6}$ used in the
$\Omega_0=0.3$ model is about 30 percent higher.
Notice that the bias in our model results solely from the difference
in counting `galaxies' and dark matter particles, and so it does not
include any {\it velocity} bias resulting from the fact 
that the velocities of galaxies
may be systematically different from those of dark matter particles.
The ratio between the observed $\sigma_{12}$ and the mean of the 
predicted ones is about 0.9 for the $\Omega_0=0.2$ model and 0.8 for the
$\Omega_0=0.3$ model. Thus, only a low level of velocity
bias between galaxies and mass is needed in these two models.  
For the $\Omega_0=1$ model, the shape of the predicted PVD is 
significantly different from that observed at small separations.
This is the case because the shape of the real-space correlation
function given by this model is significantly steeper than the 
observed one. The amplitude of $\sigma_{12}$ at $r_p=1\mpc$
predicted by this model is about 20 percent higher than the 
observed value, and so a velocity bias at the level of
0.85 will be enough to bring its prediction in agreement
with the observation. Recall that in this model the value of 
$\sigma_8\Omega_0^{0.6}$ is assumed to be 0.62 which is 
about 20 percent higher than that given by the abundance of 
clusters. Since $\sigma_{12}$ is roughly proportional to 
$\sigma_8\Omega_0^{0.5}$, we expect that the value
of $\sigma_{12}(1\mpc)$ predicted by an $\Omega_0=1$ 
model with $\sigma_8=0.5$ is at the observed level  
even in the absence of any velocity bias. 
However, the comparison between model predictions and
the observational result is more complicated in this model.
Since galaxies are positively biased relative
to the mass by a factor of about 2, the PVD of galaxies may also be
higher than that of the mass, if galaxies form at high
peaks in the mass density field (e.g. Davis et al. 1985). 
The level of velocity bias will then depend on the details
of the relationship between galaxies and density peaks. 

Figure 5 explicitly shows the spatial bias factor
$b^{\rm CW}(r_p)$ and the velocity bias factor $b^{\rm CW}_v (r_p)$ 
given by our simple (Cluster-Weighting)
bias scheme for the three cosmogonies.  
Following convention, we define $b^{\rm CW}(r_p)$
and $b^{\rm CW}_v (r_p)$ as $\sqrt{w^{\rm CW}(r_p)/w(r_p)}$ and
$\sigma_{12}^{\rm CW}(r_p)/ \sigma_{12}(r_p)$ respectively, where the
superscript `CW' labels quantities incorporating the 
Cluster-Weighting bias scheme. As expected, both the  
spatial and the velocity bias factors resulted from our 
bias scheme are scale-dependent, with stronger
(anti)biases on smaller scales.  

 Having shown that our simple phenomenological bias model may
be able to explain the discrepancy between the model predictions 
and observational results for the TPCF and PVD, we now ask whether such 
a bias is theoretically and observationally plausible. 
The bias proposed here is in the opposite sense to that proposed 
for the standard CDM model (e.g. White et al. 1987), 
because it means that the number of
galaxies formed per unit mass is suppressed rather than 
enhanced in high density regions. 
One possibility for this to happen is that galaxies
are systematically more massive, and so the number
of galaxies per unit mass is lower (although the light per unit
mass might be higher), in rich clusters than in poor clusters 
and groups. It is well known that rich clusters usually contain
big galaxies like cDs. These galaxies are about 10 times 
more massive than normal $L^*$ galaxies, and so their share in
the mass of clusters is under-represented by their number. 
However, cD type of galaxies are rare, and it is not clear
whether the luminosity functions are significantly different
between cluster and field galaxies.
Another possibility is that galaxies in clusters 
have systematically higher mass-to-light ratios
than field galaxies.  In the standard picture of galaxy formation, 
galaxies in clusters form earlier than those in the field. 
Because of their old age, these galaxies might appear fainter in 
the observational bands and so may be under-represented in a 
magnitude-limited sample. Such a phenomenon has indeed been found 
observationally (Carlberg et al. 1997b) and
in some hydro/N-body simulations (e.g. Jenkins et al. 1996).
 
The bias model invoked here is also consistent with current
observations. Based on their detailed photometric and spectroscopic
observations of rich clusters in the CNOC survey, Carlberg et
al. (1996) found that the number of galaxies [brighter than $M_r^K= -
18.5$ mag] per unit mass is systematically lower in clusters with
higher velocity dispersions. The data show that the number of galaxies
per unit mass in clusters with velocity dispersions $\ga 1000\kms$ is
lower by a factor of about 1.5 than in poorer clusters. The implied
level of bias is thus compatible with what is assumed in our bias
model. At the moment the observational data are still sparse and the
observed trend is not very significant. More observational data are
needed to test the model we are proposing here. 

It is important to emphasize that our model requires only that the
{\it number} of galaxies (brighter than $-18$ mag) per unit mass 
decreases with cluster mass. This does not necessarily mean that
the mass-to-light ratio of clusters increases with cluster
mass by a proportional amount, unless the shape of the luminosity
function of cluster galaxies is completely independent of cluster
mass. In fact, the observational data of the CNOC clusters, which
indicate a decrease of the number of galaxies per unit mass with the
cluster mass, do not show any evidence for an increase of the
mass-to-light ratio with cluster mass (Carlberg et al. 1996). 
This may already indicate that the shape of luminosity function 
does depend on cluster mass to some degree. Since the cosmic
density parameter ($\Omega_0$) derived from cluster virial masses  
is based on the mass-to-light ratio, instead of the mass-to-number
ratio, of clusters, our bias model does not necessarily imply
a lower $\Omega_0$ value than that derived by Carlberg et al.

\section{DISCUSSION}

 In this paper we have presented detailed 
comparisons with the observational results
only for three spatially-flat 
CDM models. Some of the results may,
however, apply to other interesting models of structure formation.
For open CDM models with $\Omega=0.2$-0.3, $\sigma_8\sim 1$
and $\Gamma\sim 0.2$, the two-point mass correlation functions
are slightly steeper than those for flat models
with similar model parameters (see e.g. Mo, Jing \& B\"orner 1997;
Jenkins et al. 1996; Colin, Carlberg \& Couchman 1996).
The PVD of mass on small scales is also slightly higher
(e.g. Mo, Jing \& B\"orner 1997; Colin, Carlberg \& Couchman 1996). 
Thus, the results for low-$\Omega$ open models 
are qualitatively the same as those for the low-$\Omega$ flat models 
considered in this paper, except perhaps that for them a higher value of 
$\alpha$ is needed to match the observational results. 
Another interesting case is the mixed dark matter (MDM) models
in which 20-30 percent of the cosmic mass is in massive neutrinos 
(e.g. Jing et al. 1993; Klypin et al. 1993; Jing et al. 1994; Ma \&
Bertschinger 1995; Ma 1996).
The clustering properties of dark matter predicted by such models
are similar to those given by the CDM model with $\Omega_0=1$
and $\Gamma=0.2$. For this class of models, the predicted TPCF
for the mass is quite flat and is consistent with that derived from the
angular correlation function of galaxies in the APM survey
(Jing et al. 1994; Jenkins et al. 1996). Thus, the kind of bias
we propose here may not be needed for such models to match 
the observed TPCF. The PVDs of mass predicted by these models 
are generally lower than those predicted by the standard CDM models
with the same $\sigma_8$ (Jing et al. 1994), because 
of their smaller clustering power on small scales.  
We thus expect that such models with $\sigma_8\Omega_0^{0.6}\sim 0.5$
are consistent with the observational result on PVD.

 Based on X-ray emissions and galaxy kinematics in clusters
of galaxies, Bahcall \& Lubin (1994) and Carlberg et al. (1997a) 
found little evidence for a large velocity bias in clusters of 
galaxies. The velocity bias found in (some) N-body/hydro simulations 
of galaxy formation is also modest, with 
$\sigma_{\rm gal}/\sigma_{\rm DM}\sim 0.7$-0.9 
(e.g. Katz et al. 1992; Cen \& Ostriker 1992; Frenk et al. 1996). 
The finite sizes of galaxies may reduce the peculiar velocity of
galaxies up to 10 percent (Suto \& Jing 1997).
These results imply that a large velocity bias between
galaxies and mass is not supported by current observations or 
simulations. Unfortunately, these results are still uncertain.
As we have shown, the PVD predicted by
CDM models with $\sigma_8\Omega_0^{0.6}\sim 0.4$-$0.5$ 
is compatible with the observational result even in the
absence of any velocity bias. Thus these models are consistent
with current observations. For models
with larger $\sigma_8\Omega_0^{0.6}$, a velocity bias is
needed to make models compatible with the observed PVD. 
Clearly, with a better understanding of the velocity bias,
our result on PVD can be used to put a stringent constraint 
on $\sigma_8\Omega_0^{0.6}$ in CDM models.

 We have shown, with the help of large N-body 
simulations,  that the real-space two-point correlation
function and pairwise velocity dispersion of galaxies 
can now both be measured reliably from the LCRS. 
We have carried out a detailed comparison between these observational 
results and the predictions of current CDM cosmogonies.
We have found that the observed two-point correlation function is 
significantly flatter on small scales than the mass correlation function 
predicted by current CDM models with $\sigma_8\Omega_0^{0.6}\sim 0.5$.
The observed pairwise velocity dispersion
is also lower than that of dark matter particles in these models. 
We have proposed a simple bias model to explain these discrepancies. 
This model assumes that the number of galaxies per unit dark 
matter mass, $N/M$, decreases with the mass of dark haloes increasing.
The predictions of CDM models with 
$\sigma_8\Omega_0^{0.6}\sim 0.4$-0.5 
and $\Gamma \sim 0.2$ are in agreement with the observational results, 
if the trend of $N/M$ with $M$ is at the level observed for rich 
clusters of galaxies. A velocity bias is needed for models
with larger $\sigma_8\Omega_0^{0.6}$.
Therefore current CDM models with $\Omega_0 h\sim 0.2$
and with cluster-abundance normalizations are consistent with 
the observed correlation function and pairwise velocity dispersion 
of galaxies. A high level of velocity bias is not required
in these models. The observational data can put
a stringent constraint on the value of $\sigma_8\Omega_0^{0.6}$
in CDM models once the level of velocity bias is known.

\acknowledgments 

We are grateful to Yasushi Suto for helpful discussions, and to Simon
White for a careful reading of the manuscript. We thank the LCRS group
for making their data publicly available, Douglas Tucker for
helpful information on the catalog, and the referee of the paper for a 
detailed report. Y.P.J. acknowledges the receipt of
a JSPS Postdoctoral Fellowship. The research was in part supported by
the Grants-in-Aid of the Ministry of Education, Science, Sports and
Culture of Japan ((No.07CE2002, No.96183). The simulations were
carried out on VPP/16R and VX/4R at the Astronomical Data Analysis
Center of the National Astronomical Observatory, Japan.

\appendix

\section{Testing the statistical methods}

In this appendix we test whether the statistical methods
described in Section 3.1 can yield unbiased measurements of the 
TPCF and the PVD from a sample like the LCRS. Here we generate 
20 mock catalogs from one simulation of the $\Omega_0=0.2$ model, 
{\it without including} the fiber collision effect. 
Fiber collisions will cause some bias in the statistics, 
as will be discussed in Appendix B.

The projected TPCF measured directly from the mock samples
is plotted as the solid circles in Figure 6, with the error bars 
representing the standard deviation of the mean. 
The thick solid curve shows $w(r_p)$ given by the 
real-space two-point correlation function $\xi(r)$: 
\beq\label{A1}
w(r_p)=\int_0^\infty \xi(\sqrt{r_p^2+\pi^2}) d\pi,
\eeq 
where we have used the empirical model as described in 
Mo, Jing \& B\"orner (1997; see also Peacock \& Dodds 1996) to 
calculate the model prediction for $\xi$. 
This $w(r_p)$ is what we want to recover from the mock samples. 
As shown in the figure, the result derived 
from the mock samples agrees well with that given by the empirical
model. (The slightly higher value given by the mock samples on scales 
larger than $10\mpc$ is due to the fact that the particular 
realization of the simulation used happens to have systematically higher 
large-scale power than average). Thus, the statistical method 
we have used can give an unbiased estimate of the projected TPCF.
The good agreement between the mock $w(r_p)$ and that given by the
empirical formula at the small scales ($r_p\approx 0.1\mpc$) indicates that the
simulations used have sufficient resolution for the purpose of this paper.

The measurement of the PVD from mock samples based on the redshift
distortion of the TPCF (as described in Section 3.1) is compared in
Figure 7 to the {\it true} PVD which we want to measure. The true PVD
is defined as $\langle \left\{[{\bf v}_{12}({ r})-\langle {\bf
    v}_{12}({ r})\rangle]/3\right\}^2\rangle ^{1/2}$ where ${\bf
  v}_{12}({r})$ is the 3-D relative velocity of two particles with
separation $r$ and $\langle \cdot\cdot\cdot \rangle$ denotes
average over all pairs at separation $r$. The true PVD can be
estimated from the three-dimensional velocities of particles in the
simulation. Two infall models, the self-similar solution (equation
\ref{infall}) and the infall pattern obtained directly from the
simulation, are used for reconstructing the PVD from the mock samples.
The two models give very similar results on scales
$r_p<10\mpc$. Comparing the PVD reconstructed from the redshift
distortion with the true value, we see that the two agree with each
other qualitatively.  However, the difference between the two
quantities is quite significant even if the real infall pattern is
used. This difference arises from the fact that the distribution of
the pairwise velocities is not perfectly exponential and that the true
PVD depends significantly on the separations of pairs in real
space.  Thus, to make a rigorous comparison between models and
observational results, these systematics must be treated 
very carefully.

\section{Quantifying the effect of fiber collision}

In this Appendix, we apply the same test as in Appendix A 
to mock samples which {\it include} the fiber collision effect. 
The purpose of this is to quantify the effect of fiber collisions 
on the two statistics presented in this paper. 
Figure 8 shows the ratio of $w(r_p)$ measured from the 20
mock samples with and without fiber collisions. 
As expected, the effect of the fiber collisions is larger on smaller
scales. The value of $w(r_p)$ is reduced by
up to 15\% on the smallest scale, $r_p \sim 0.1\mpc$. 
On scales $r_p\ga 2\mpc$, the effect is very small. The scatter 
of the ratio among different mock samples is very small, 
which means that the fiber collision effect is
systematic and can be corrected easily in the analysis. 
The thin line in Fig.6 shows $w(r_p)$ measured from the 20 mock 
samples with fiber collisions included.

The effect of the fiber collisions on the determination of PVD
is shown in Fig.9. Here we plot the difference in
$\sigma_{12}(r_p)$ estimated by comparing
20 mock samples
in which fiber collisions are included with 20 others 
where fiber collisions are excluded.
Fiber collisions reduce the value of $\sigma_{12}(r_p)$ 
by about $20\kms$ for $r_p\la 3\mpc$ and by 
a smaller amount for larger $r_p$. 
This result depends very weakly on the infall model adopted.

We have tested the robustness of the above results to the
change in the intrinsic clustering,
applying the same tests to the cluster
weighted mock samples (section 3.2) of the $\Omega_0=0.2$ model.
The statistical results are the same as those presented in Figures 8
and 9. Thus the effects of fiber collisions on TPCF and PVD are
not sensitive to the intrinsic clustering, and the results presented
in this Appendix can be used to correct for the fiber collision effects 
in the LCRS.

\vfill
\eject

\begin{figure}
\epsscale{1.0} \plotone{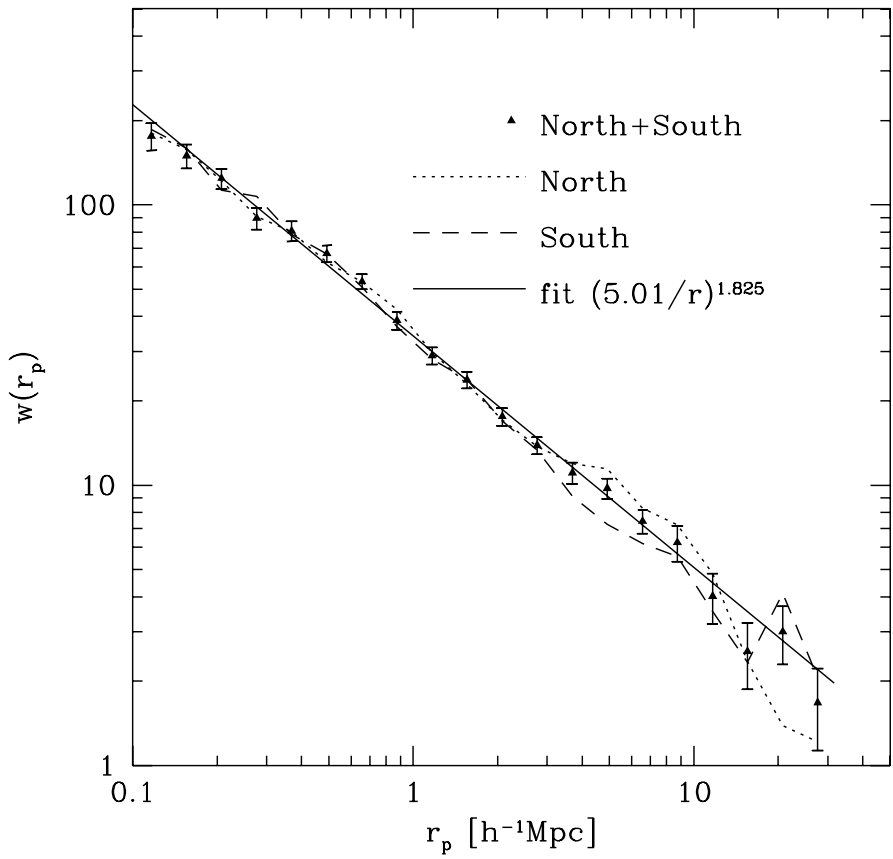}
\caption{
  The projected two-point correlation function measured from the Las
  Campanas Redshift Survey (filled triangles). Error bars are
  $1\sigma$ deviations given by bootstrap resampling. 
  Results for the north and south subsamples are
  shown by the dotted and dashed lines, respectively.  
  The solid line
  is the power-law fit to the correlation function
  of the total sample.  
}\end{figure}

\begin{figure}
\epsscale{1.0} \plotone{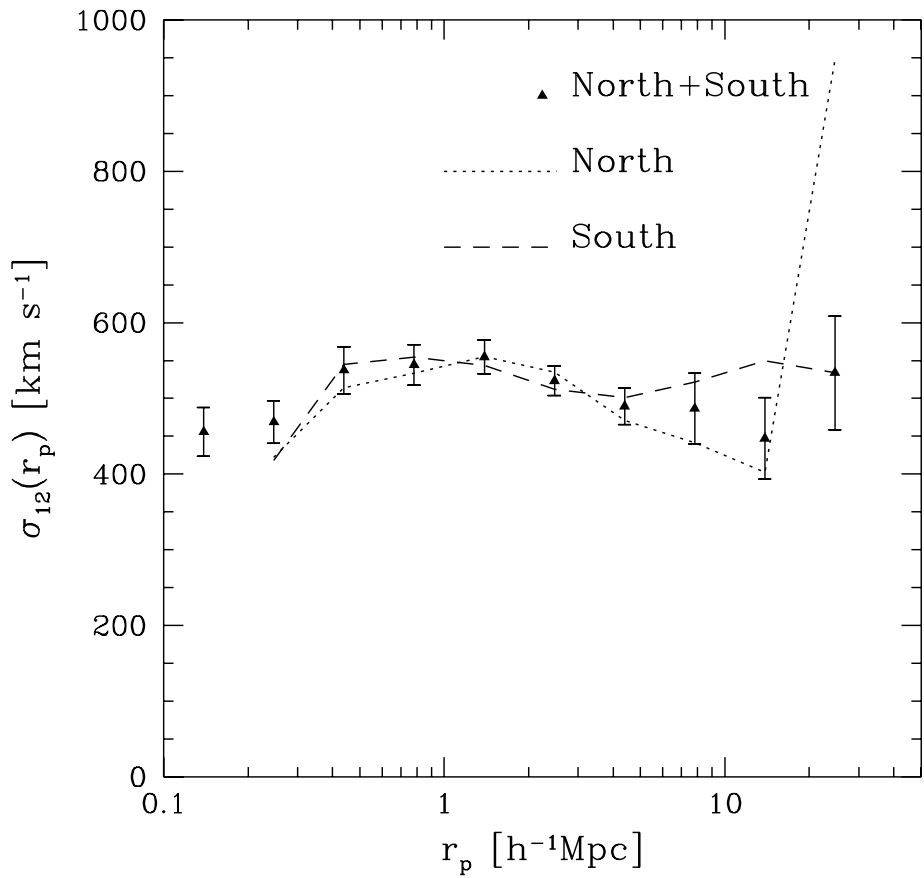}
\caption{
  The pairwise velocity dispersion $\sigma_{12}(r_p)$ measured from
  the Las Campanas Redshift Survey (filled triangles). Error bars are
  $1\sigma$ deviations given by bootstrap resampling. 
  Results for the north and south
  subsamples are shown by the dotted and dashed lines, respectively.}
\end{figure}

\begin{figure}
\epsscale{0.8} \plotone{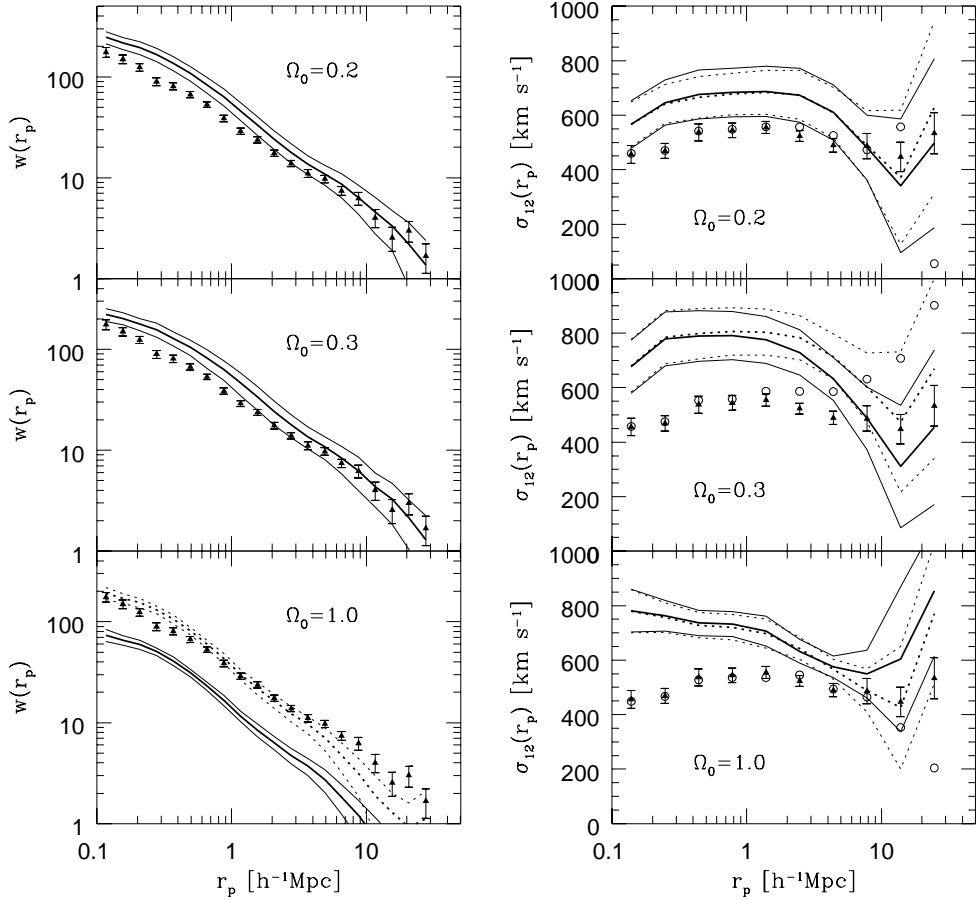}
\caption{Comparison of the predictions of CDM models with the LCRS 
  results. {\it Left panels}-- the projected two-point correlation
  function. Triangles show the observational result from Fig. 1. 
  The mean value and the $1\sigma$ limits predicted by 
  the CDM models are shown by the thick and thin lines respectively.
 The dashed lines in the lowest panel
 are obtained by shifting the solid lines upwards by an amount
 of $1/\sigma_8^2$.
  {\it Right panels}-- the pairwise velocity dispersion. 
  Thick and thin lines show the mean value and the 1$\sigma$ limits 
  predicted by the CDM models. The solid lines show the results
  obtained from the self-similar infall model, while the dashed lines 
  are those obtained from the real infall pattern given by 
  simulations of the model under consideration. 
  Triangles show the result for the LCRS obtained from
  the self-similar infall model; circles show the result
  when the infall pattern given by the simulations is used. 
  Error bars for the LCRS results are shown only for the 
  self-similar infall model.
}\end{figure}

\begin{figure}
\epsscale{1.0} \plotone{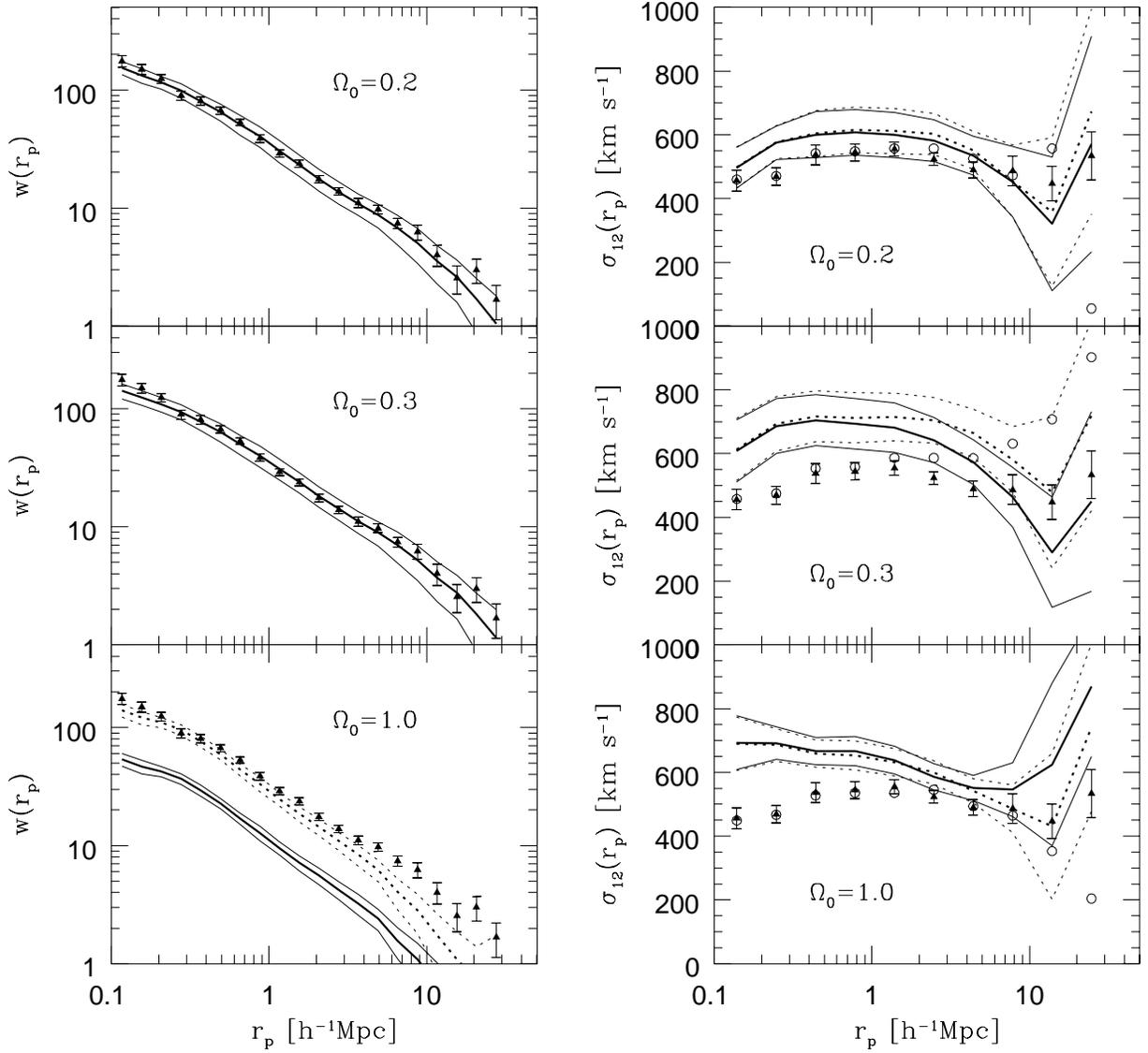}
\caption{The predictions of the three CDM models incorporating
 a simple bias model (see text). The lines and symbols have the same
 meaning as in Fig.3. The dashed lines in the lower-left panel
 are obtained by shifting the solid lines upwards by an amount, 
 $1/\sigma_8^2$. 
}\end{figure}

\begin{figure}
\epsscale{1.0} \plotone{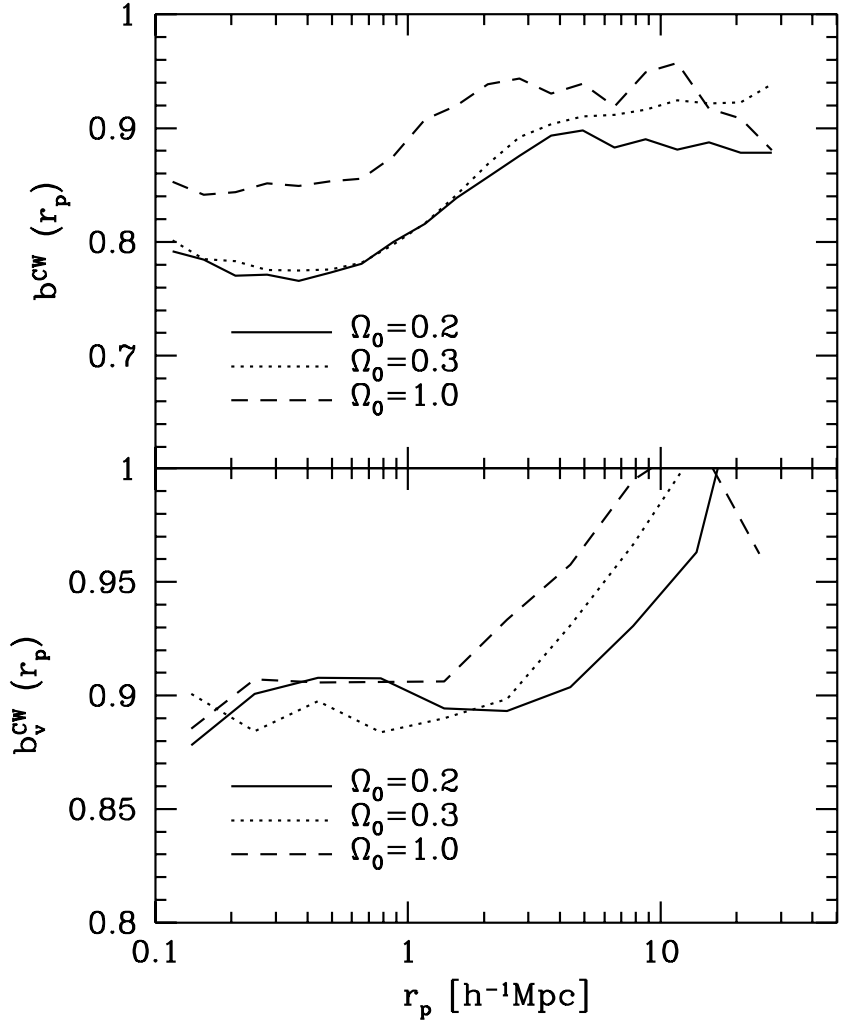}
\caption{The biases predicted by the bias model on the projected TPCF
  ({\it upper panel}) and the PVD ({\it lower panel}) for the three
cosmological models. 
}\end{figure}

\begin{figure}
\epsscale{1.0} \plotone{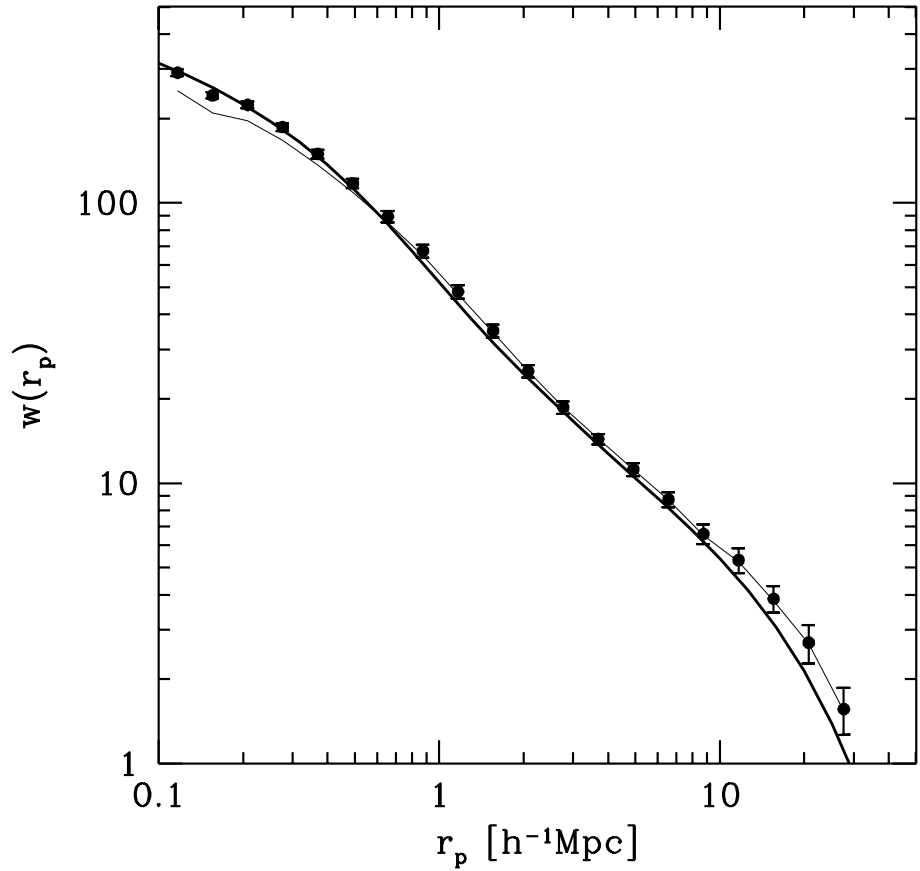}
\caption{
  The projected two-point correlation function for the
  $\Omega_0=0.2$ model estimated from 20 mock
  samples without fiber collisions (filled circles). 
  Error bars are the ($1\sigma$) standard deviations of the mean 
  from the mock samples. The thick solid line is the model prediction
  based on the empirical fitting formula. The thin
  line shows the result for the 20 mock samples including
  the fiber-collision effect.  
}\end{figure}

\begin{figure}
\epsscale{1.0} \plotone{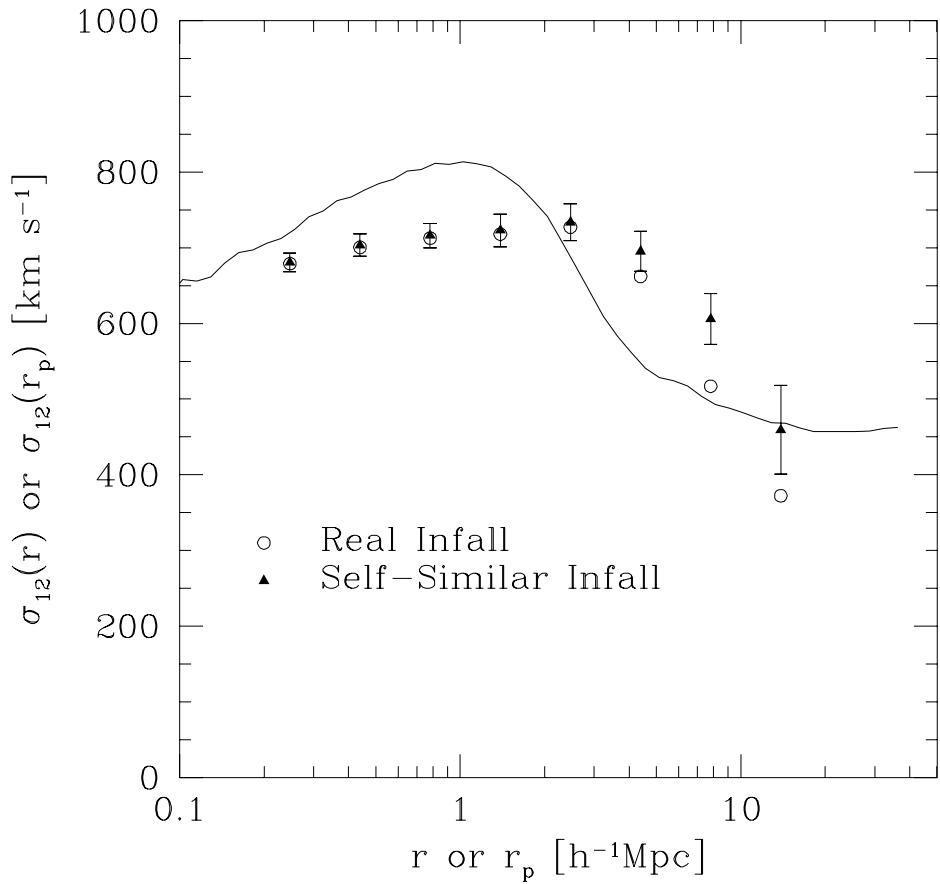}
\caption{
  The pairwise velocity dispersion $\sigma_{12}(r_p)$ measured from
  the redshift distortion of the two-point correlation function of the
  20 mock samples without fiber collisions. Two infall models are
  adopted for $\overline {v_{12}}(r)$: the self-similar infall model
  (triangles) and the infall pattern derived directly from the 
  simulations (circles).  The {\it true} pairwise velocity
  dispersion given by the 3-dimensional velocities in the simulations 
  is shown as the solid line.
  Error bars are the ($1\sigma$) standard deviations of the mean 
  from the mock samples.
}\end{figure}

\begin{figure}
\epsscale{1.0} \plotone{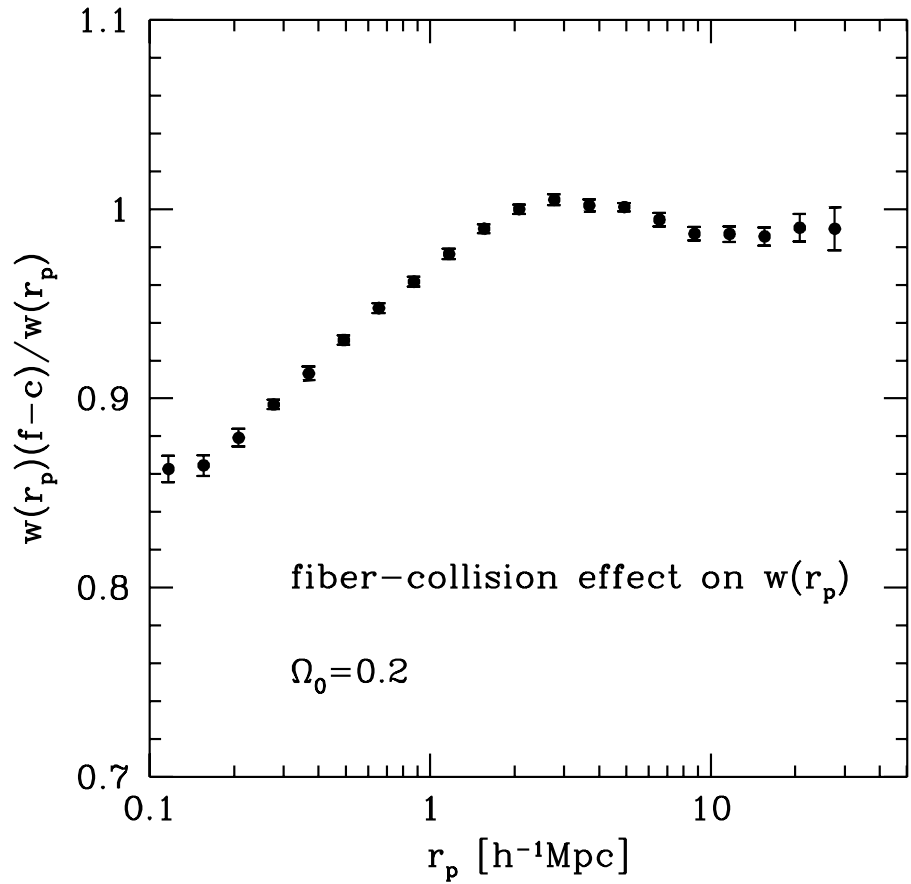}
\caption{
  The ratio of the projected two-point correlation functions measured
  from 20 mock samples with and without fiber collisions.
  Error bars are the ($1\sigma$) standard deviation of the mean from 
  the mock samples.
}\end{figure}

\begin{figure}
\epsscale{1.0} \plotone{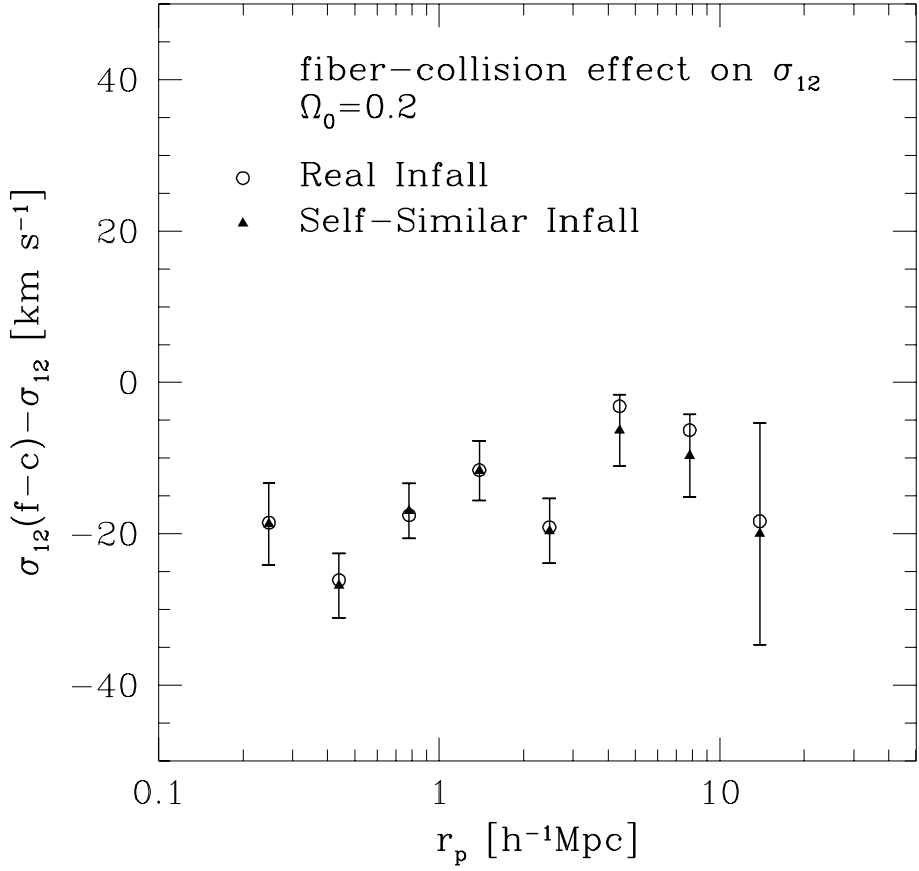}
\caption{
The difference between the pairwise velocity dispersions 
estimated for 20
mock samples with and without fiber collisions. 
Results are shown for the two infall models, as in Fig.6.
Error bars are the ($1\sigma$) standard deviation of the mean from 
the mock samples.
}\end{figure}

\end{document}